\documentclass[aps,prl,twocolumn,floatfix,superscriptaddress,nobalancelastpage]{revtex4-1}
\usepackage[english]{babel}
\usepackage{amsfonts}
\usepackage{amsmath}
\usepackage{graphicx}
\usepackage[caption=false]{subfig}

\usepackage{soul}
\usepackage{color}

\newcommand{\expo}[1]{\mathrm{exp}\left(#1\right)}

\newcommand{\cose}[1]{\mathrm{cos}\left(#1\right)}

\newcommand{\seno}[1]{\mathrm{sin}\left(#1\right)}

\newcommand{\daga}[1]{#1^{\dagger}}
\newcommand{\ex}[1]{\:e^{#1}}
\newcommand{\Ket}[1]{\left|#1\right>}

\newcommand{\KetBra}[2]{|#1\rangle\langle#2|}
\newcommand{\Nc}{\mathcal{N}}
\newcommand{\jota}[1]{J_{#1}\left(\frac{\Delta}{\omega_D}\right)}

\begin{document}

\title{Driving-induced amplification of non-Markovianity in open quantum systems evolution }

%

\author{P. M. Poggi}
\email{ppoggi@df.uba.ar}
\affiliation{Departamento de F\'{\i}sica Juan Jose Giambiagi 
 and IFIBA CONICET-UBA,
 Facultad de Ciencias Exactas y Naturales, Ciudad Universitaria, 
 Pabell\'on 1, 1428 Buenos Aires, Argentina}

\author{F. C. Lombardo}
\affiliation{Departamento de F\'{\i}sica Juan Jose Giambiagi 
 and IFIBA CONICET-UBA,
 Facultad de Ciencias Exactas y Naturales, Ciudad Universitaria, 
 Pabell\'on 1, 1428 Buenos Aires, Argentina}

\author{D. A. Wisniacki}
\affiliation{Departamento de F\'{\i}sica Juan Jose Giambiagi 
 and IFIBA CONICET-UBA,
 Facultad de Ciencias Exactas y Naturales, Ciudad Universitaria, 
 Pabell\'on 1, 1428 Buenos Aires, Argentina}
\date{\today}


\begin{abstract}
	Non-Markovian effects arising in open quantum systems evolution have been a subject of increasing interest over the past decade. One of the most appealing features of non-Markovianity (NM) is that it captures scenarios where loss of information and coherence are reversible, and thus a temporary backflow of information from the environment to the system is possible. In this work we study the interplay between the degree of non-Markovianity and the action of time-dependent control fields in an open two-level quantum system. We find that periodical modulation of a field acting solely on the system can greatly enhance the degree of non-Markovianity with respect to the undriven case. We show that this effect is present only when the coupling between system and environment is weak. Remarkably, the enhancement disappears at strong coupling, which is usually the regime where non-Markovian effects are expected to be more pronounced.
\end{abstract}


\maketitle

\section{Introduction}
How does the presence of a driving field affect the non-Markovian features of open quantum system dynamics? This question arises naturally when we consider two converging scenarios: first, the enormous advances made in the past decades in using electric, magnetic and optical fields to precisely control nano- and subnano-scale physical systems, from single site optical manipulation in ion traps \cite{bib:wineland2015,bib:blatt2016}, to coupling single electron to photons in silicon quantum dots \cite{bib:petta2016} and the realization of long-lived coherent qubits in superconducting circuits \cite{bib:devoret2016}. And second, the great interest raised during the past decade over non-Markovian dynamics, which refers to the evolution of open quantum systems which show a substantial deviation from the usual Markovian scheme \cite{bib:breuer2016,bib:devega2017}.\\

Over the past decade, numerous works have studied the role of NM in a variety of scenarios in open quantum system evolution. However, the debate about whether NM is a useful resource for quantum technologies such as computing, simulation and communication \cite{bib:maniscalco2014} is far from being settled. For instance, studies which assessed the effectiveness of optimal control methods \cite{bib:rabitz1998,bib:krotov} in open quantum system evolutions showed that NM allowed for an improved controllability \cite{bib:schmidt2011,bib:koch2015,bib:pachon2016}. However, non-Markovian effects were also related to reduction of efficiency in dynamical decoupling schemes \cite{bib:addis2015}. In a similar direction, the task of generating steady entangled states in open quantum systems has been showed to rely upon non-Markovian effects in some cases \cite{bib:huelga2012}, but not in others \cite{bib:cormick2013}, where NM does act as catalyst to speed up the process. Finally, the problem of the maximum speed of evolution for open quantum systems, usually termed quantum speed limit (QSL) \cite{bib:delcampo2013}, has also been linked with NM. In particular, it was argued that non-Markovian effects could actually speed up the system evolution and thus reduce the QSL time \cite{bib:deffner2013,bib:meng2015}. A recent work on the subject revealed that non-Markovian effects could not be inferred from the QSL bounds \cite{bib:mirkin2016}.\\

In this work, we tackle the question that opens this paper. In particular, we investigate to what extent external driving acting solely on the system can increase NM with respect to the undriven case. To this end, we consider a two-level system described by a time-periodic Hamiltonian interacting with a structured environment. We find that the driving has a peculiar effect on the non-Markovian character of the system dynamics: it can generate a large enhancement of the degree of NM with respect to the static case, but only when the coupling between system and environment is weak. For strong coupling, where non-Markovian effects are usually more prominent, the driving cannot increase the NM substantially. We finally show analytically that, for the case of high-frequency driving, the system tends to decouple from the environment. This effect provides insight about the inability of the driving to enhance the degree of NM in that regime.   \\



\section{Model and methods} \label{sec:model}


Let us consider a two-level quantum system interacting with a bath of harmonic oscillators. The total Hamiltonian which describes this model reads (we set $\hbar=1$ from here on)
\begin{equation}
H = \bar{\omega}_0(t)\sigma_+\sigma_- + \sigma_x\sum_k g_k(b_k+b_k^\dagger) + \sum_k\bar{\omega}_k b_k^\dagger b_k,
\label{ec:hami}
\end{equation}
\noindent where $\sigma_\pm=\sigma_x\pm i\sigma_y$ and $\sigma_\alpha$ with $\alpha=x,y,z$ are the Pauli matrices, $b_k$ and $b_k^\dagger$ are the usual annihilation and creation operators corresponding to the $k$th mode of the bath, $g_k$ are the coupling constants and $\bar{\omega}_0(t)$ is the time dependent energy difference between states $\Ket{0}$ and $\Ket{1}$ of the two-level system, which we will assume to be of the form
\begin{equation}
  \bar{\omega}_0(t) = \bar{\Omega}_0 + \bar{\Delta}\cose{\bar{\omega}_D t}.
  \label{ec:omega_t}
\end{equation}

All of the information of the environment relevant to the evolution of the reduced density matrix of the two-level system is condensed in the bath time-correlation function

\begin{equation}
  C(t-t')=\mathrm{tr}\left[\rho_B B(t)B(t')\right]=\int_0^{\infty} d\omega\: J(\omega)\ex{-i\omega(t-t')},
  \label{ec:correlat}
\end{equation}

\noindent where $B(t)=\sum_k g_k(b_k \ex{-i\bar{\omega}_k t}+\daga{b_k} \ex{i\bar{\omega}_k t})$, $\rho_B$ is the initial state of the environmental modes and $J(\omega)$ denotes its spectral density. Here we consider $J(\omega)$ to be of Lorentzian form, i.e.
\begin{equation}
  J(\omega)=\frac{\bar{\gamma}_0}{2\pi}\frac{\lambda^2}{(\omega-\bar{\Omega}_0)^2+\lambda^2)},
  \label{ec:spectral}
\end{equation}
\noindent where  $\bar{\Omega}_0$ is a characteristic frequency, $\lambda$ determines the broadening of the spectral peak, and $\bar{\gamma}_0$ sets the coupling strength between the system and the bath. A known example of a physical system which is accurately described by this model is that of an atom coupled to an imperfect cavity \cite{bib:petruccione}. We point out that, by considering this model, we are studying an structured environment which is usually the scenario where non-Markovian effects are more prone to manifest themselves \cite{bib:devega2017}. By inserting expression (\ref{ec:spectral}) in (\ref{ec:correlat}) we can evaluate the correlation function which yields, when $\bar{\Omega}_0\gg\lambda$,
\begin{equation}
C(t-t')=\frac{\bar{\gamma}_0\lambda}{2}\ex{-(\lambda+i\bar{\Omega}_0)(t-t')}.
\label{ec:correlat2}
\end{equation}

If we now assume the validity of the rotating wave approximation (RWA) and drop counter-rotating terms from the interaction Hamiltonian, we can rewrite the approximated Hamiltonian as 
\begin{equation}
  H = \bar{\omega}_0(t)\sigma_+\sigma_- + \sum_k g_k\left(\sigma_+ b_k+\sigma_- b_k^\dagger\right) + \sum_k\bar{\omega}_k b_k^\dagger b_k,
  \label{ec:hami_rwa}
\end{equation}

By following the procedure of Ref. \cite{bib:petruccione}, we can obtain the exact equations of motion to all orders in the system-environment coupling. Due to the conservation of the total excitation number and assuming the environment to be initially prepared in the vacuum state, the dynamics of the total system takes place solely in the single-excitation subspace. This allows us to characterize the dynamics of system through a single complex-valued function $G(t)$. The reduced density matrix elements of the two-level system in the $\left\{\Ket{0},\Ket{1}\right\}$ basis then read

\begin{equation}
\Bigg\{ \begin{array}{r c l}
		\rho_{00}(t) & = & \rho_{00}(0)|G(t)|^2 \\
		\rho_{01}(t) & = & \rho_{01}(0)G(t)\ex{-i\epsilon(t)}
	\end{array}
\label{ec:rho_t}
\end{equation}

\noindent where $\epsilon(t)=\bar{\Omega}_0 t + \left(\bar{\Delta}/\bar{\omega_D}\right)\seno{\bar{\omega}_D t}$.  The equation for $G(t)$ can be casted in a time-local form due to the exponential dependence of $C(t-t')$. By defining dimensionless parameters variables $\tau\equiv\lambda t$ and $x=\bar{x}/\lambda$, where $x$ is any parameter with units of energy ($\bar{\Delta}$ or $\bar{\gamma}_0$ for instance), the dynamical equation for $G(t)$ is
\begin{equation}
G''(\tau)+\left[1-i\Delta\mathrm{cos}\left(\omega_D\tau\right)\right]G'(\tau)+\frac{\gamma_0}{2}G(\tau)=0,
\label{ec:ec_rwa}
\end{equation}
\noindent where we established the notation $f'(\tau)\equiv\frac{df}{d\tau}$. Equation (\ref{ec:ec_rwa}) is the exact equation of motion for the driven two-level system coupled to a Lorentzian bath at zero temperature, under the RWA. Note that the equation also holds when $\omega_D=0$, that is, when the system Hamiltonian is time-independent, and an exact solution for $G(t)$ is available \cite{bib:petruccione,bib:pineda2016,bib:mirkin2016}. In that case, $\Delta$ plays the role of a static detuning between the system and the central frequency of the environment.\\

When we consider the system parameter $\omega_0(t)$ to be periodically driven between $\Omega_0-\Delta$ and $\Omega_0+\Delta$ at a rate $\omega_D$, then the RWA may not be well justified, specially in the high driving frequency limit. Moreover, even in the regime where the RWA describes accurately the system evolution, the non-Markovianity measures calculated with and without the RWA have been shown to produce very different results \cite{bib:makela2013,bib:sun2015}. In order to describe the reduced dynamics of the system taking into account the counter-rotating terms, we use the hierarchy equations method developed by Tanimura et al. \cite{bib:tanimura2006} and then successfully applied to a variety of interesting problems \cite{bib:nori2012,bib:sun2015,bib:chen2015,bib:guo2014,bib:sun2016}. This method can be used if (i) the initial state of the system plus bath is separable, (ii) the interaction Hamiltonian is bilinear (i.e. $H_{SB}=S\otimes B$), and (iii) if the environmental correlation function can be casted in multi-exponential form. All of these requirements are met in our system, see equations (\ref{ec:hami}) and (\ref{ec:correlat2}). We give some details about this method in the Supplementary Material.\\

We are interested in studying non-Markovian effects on the reduced dynamics of the two-level system. One of the most widespread witnesses of non-Markovianity in open quantum system evolutions was proposed by Breuer, Laine and Piillo (BLP) in Ref. \cite{bib:breuer2009}. Conceptually, the BLP criterion is inspired on the contraction of classical probability space under Markovian stochastic processes \cite{bib:breuer2016,bib:pineda2016}. For open quantum systems, it is based on the notion of distinguishability of quantum states, as quantified by the trace distance between them
\begin{equation}
D(\rho_1,\rho_2)=\frac{1}{2}||\rho_1-\rho_2||,\ \mathrm{with}\ ||A||=\mathrm{tr}(\sqrt{A^{\dagger}A}).
\label{ec:trace_dist}
\end{equation}

The BLP criterion states that a map $\Lambda_t$, where $\rho(t)=\Lambda_t[\rho(0)]$, is non-Markovian if there exists at least a pair of initial states $\rho_1(0)$, $\rho_2(0)$ such that
\begin{equation}
\sigma(\rho_1(0),\rho_2(0),t)\equiv\frac{d}{dt}D(\rho_1(t),\rho_2(t))>0
\end{equation}
\noindent for some time interval. Note that $\sigma>0$ implies that the states $\rho_1(t)$ and $\rho_2(t)$ are moving away from each other in state space, thus increasing their distinguishability \cite{bib:breuer2009}. This means that information about the initial state of the system, is re-gained: information has flowed from the environment back to the system. In Markovian dynamics, the opposite occurs: since $\sigma\leq 0$ for all times, information is continuously lost to the environment.\\

The BLP criterion can be naturally extended to define a measure of the degree of NM in a quantum process. The original proposal aims to quantify the total amount of information backflow during the evolution of the system, and it is calculated via
\begin{equation}
\Nc_{BLP}=\max_{\left\{\rho_1(0),\rho_2(0)\right\}} \int_{\sigma>0} \sigma(\rho_1(0),\rho_2(0),t')dt',
\label{ec:n_blp}
\end{equation} 
\noindent where an optimization has to be made for all pair of possible initial states. For the system we consider in this work, it is possible to show (under the RWA) that $\Nc_{BLP}$ takes it maximum value when choosing $\rho_1(0)=\KetBra{+_x}{+_x}$ and $\rho_2(0)=\KetBra{-_x}{-_x}$, i.e. the eigenstates of operator $\sigma_x$ \cite{bib:breuer2016,bib:wissmann2012}.\\

This proposal for a NM measure possesses a number of inconvenient properties, which have been pointed out and thoroughly studied by Pineda and coworkers in \cite{bib:pineda2016}. The main issue is that it overestimates the weight of fluctuations in $D(\rho_1(t),\rho_2(t))$. Note that fluctuations may arise from finite-sized environments, but also from temporal driving and the contribution of counter-rotating terms in the Hamiltonian \cite{bib:makela2013}, as we shall see later on. In Ref. \cite{bib:pineda2016}, the authors propose another method to quantify non-Markovian effects. Applied to the BLP criterion of distinguishability, this new measure is calculated as the largest revival of $D(\rho_1(t),\rho_2(t))$ with respect to its minimum value prior to the revival, i.e.
\begin{equation}
\Nc_{LR}=\max_{tf,t\leq tf}\left[D(\rho_1(t_f),\rho_2(t_f))-D(\rho_1(t),\rho_2(t))\right].
\label{ec:n_lr}
\end{equation}
This quantity clearly avoids the aforementioned pitfalls of the BLP measure as it insensitive to fluctuations by construction.

\section{Results} \label{sec:results}

\begin{figure}
	\includegraphics[width=\linewidth]{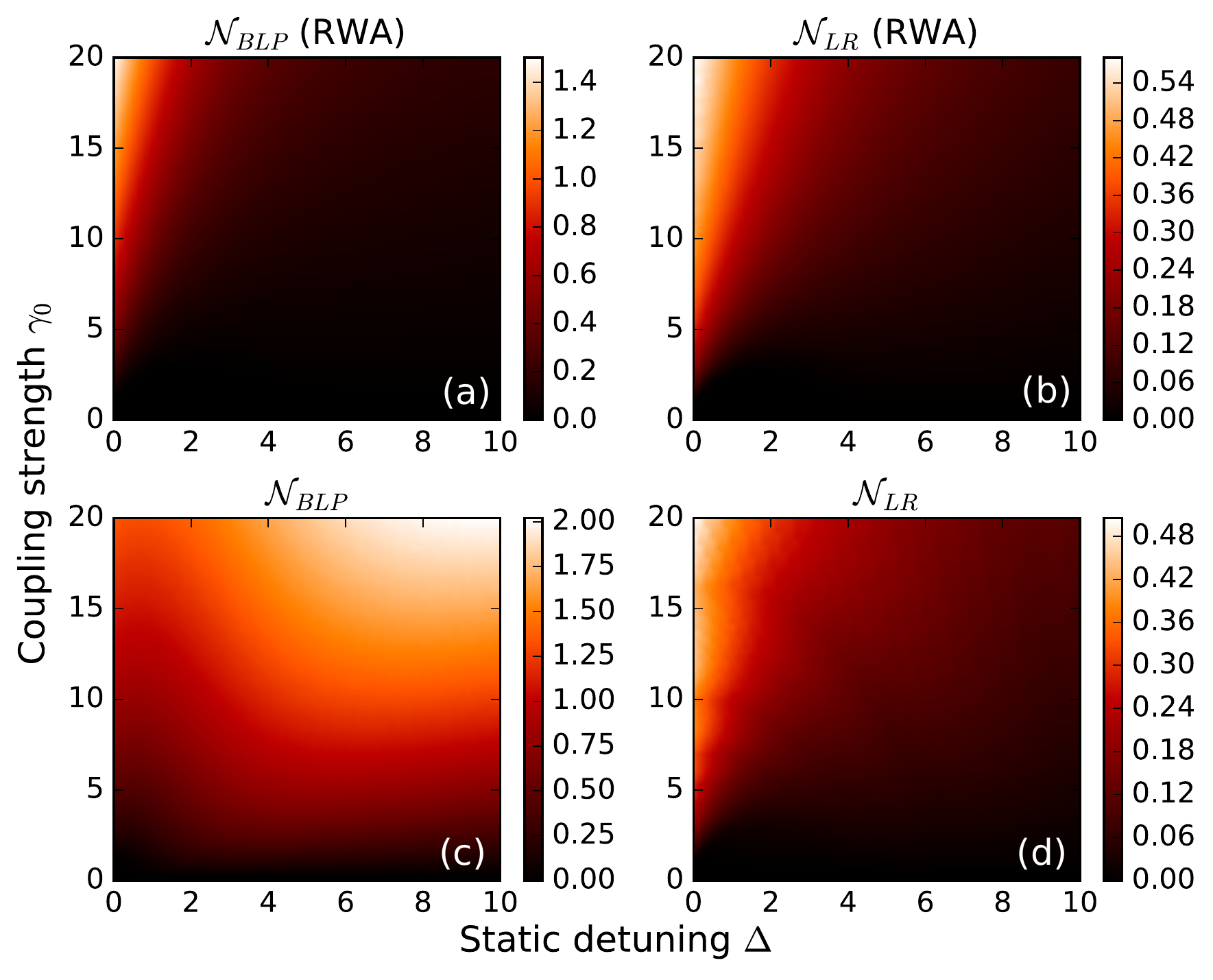}
	\caption{\label{fig:fig1} (a)-(b) Density plots for the non-Markovianity measures $\Nc_{BLP}$ and $\Nc_{LR}$ as a function of the dimensionless coupling parameter $\gamma_0$ and detuning $\Delta$ for the static case ($\omega_D=0$). Results were obtained under using the equation of motion (\ref{ec:ec_rwa}), i.e. using the RWA. (c)-(d) Same as (a) and (b), but results obtained using the hierarchy equation method (beyond RWA). Dimensionless parameter values: $\Omega_0=20$.} 
\end{figure}

Let us first consider the static case, i.e. when the driving frequency is set to $\omega_D=0$. Under the RWA, this model admits an exact solution of the equation of motion (\ref{ec:ec_rwa}) \cite{bib:breuer2016,bib:mirkin2016}. The behaviour of the NM measure $\Nc_{BLP}$ has been extensively studied in the literature \cite{bib:breuer2009,bib:rivas2014}, and it has also been compared with the largest revival measure $\Nc_{LR}$ \cite{bib:pineda2016}. In Fig. \ref{fig:fig1} (a) and (b) we plot both quantities as a function of the dimensionless coupling strength $\gamma_0$ and static detuning $\Delta$. In both cases, the NM increases with $\gamma_0$ and in general decreases with $\Delta$ for fixed $\gamma_0$ (except at weak coupling).\\

An important result arises when comparing the NM measures calculated beyond the RWA. In Fig. \ref{fig:fig1} (c) and (d) we plot $\Nc_{BLP}$ and $\Nc_{LR}$ calculated using the hierarchy equation method described in the previous section, which takes into account the effects of the counter-rotating terms in the interaction Hamiltonian. There, it can be seen that the largest revival measure $\Nc_{LR}$ shows a very similar behaviour compared to the RWA case, while the BLP measure $\Nc_{BLP}$ not only takes larger values \cite{bib:makela2013} but also displays a completely different dependance with the system parameters. Note that, by setting $\Omega_0>>1$ (i.e. $\bar{\Omega}_0\gg\lambda$), we are working in a regime where the RWA and non-RWA evolutions are very similar, apart from fluctuations. This results shows us that the largest revival measure is much more reliable than the BLP measure, which leads to an incorrect quantification of non-Markovian effects in the presence of fluctuations. This robustness is a desired characteristic for a NM quantifier, specially when we address the system to be driven by a time-periodic field, as we shall do next.\\


We now turn our attention to the time-dependent case, by tuning the driving frequency $\omega_D$ to non-zero values. In order to assess quantitatively how the non-Markovian character of the evolution is affected by the driving, we define a figure of merit called ``relative NM value'' as

\begin{equation}
\Nc_{x}^{rel}(\gamma_0,\Delta,\omega_D)=\frac{\Nc_x(\gamma_0,\Delta,\omega_D)}{\max\limits_{\Delta'}\ \Nc_x(\gamma_0,\Delta',0)},
\label{ec:fig_merit}
\end{equation}

\noindent where $x=\mathrm{BLP}$ or $\mathrm{LR}$. This quantity measures the obtained NM value for the driven case in units of the maximum possible degree of NM achieved in the system with no driving ($\omega_D=0$), at a fixed value of the coupling parameter $\gamma_0$. As such, it allows us to assess how the NM is affected by the periodic driving itself and not by, for example, introducing new interactions or injecting energy in the system Hamiltonian (which has been studied, for example, in \cite{bib:addis2015,bib:haikka2011}). This is because the driven system can be interpreted as the static Hamiltonian
\begin{equation}
H_s=(\Omega_0+\Delta)\sigma_+\sigma_-,
\end{equation}
\noindent but with a time-dependent detuning $\Delta\rightarrow\Delta\:\mathrm{cos}(\omega_D\tau)$. In other words, by studying $\Nc_{x}^{rel}$ we are analyzing the effects of dynamically changing the detuning in time, and not merely by increasing or diminishing its value.\\

Our main interest is to see to what extent the external driving, which acts only on the two-level system, can enhance the non-Markovian character of the evolution. To do this, we compute the maximum of $\Nc_{x}^{rel}$ over all values of the driving frequency and amplitude
\begin{equation}
M_{x}(\gamma_0)=\max_{\Delta,\:\omega_D}\ \Nc_{x}^{rel}(\gamma_0,\Delta,\omega_D),
\label{ec:Mmax}
\end{equation}

\noindent where $x=\mathrm{BLP}$ or $\mathrm{LR}$. $M_{x}$ indicates the maximum relative NM value, and so quantifies how much bigger can the NM measure be in the driven case when compared to the static case. Of course, $M_x\geq 1$ since, even in the worst case, the driven case will be equal to the static case for $\omega_D=0$. \\
 
\begin{figure}[!t]
	\includegraphics[width=\linewidth]{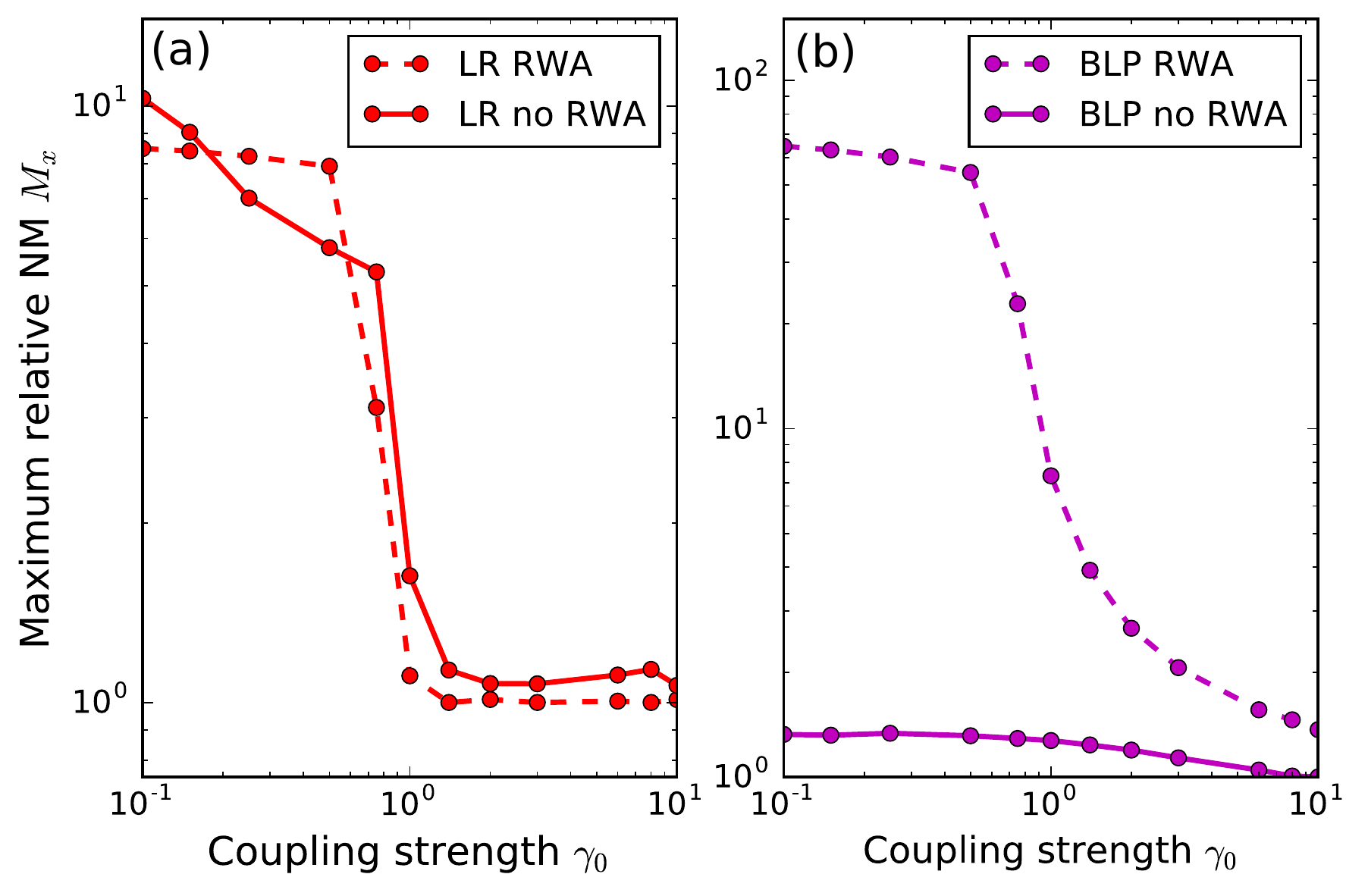}
	\caption{\label{fig:fig2} Maximum relative NM value, as defined in Eq. (\ref{ec:Mmax}), as a function of the coupling parameter $\gamma_0$, calculated with (dashed lines) and without (full lines) the rotating wave approximation. Values were obtained for (a) LR measure, (b) BLP measure. Dimensionless parameter values: $\Omega_0=20$, $\Delta,\:\omega_D  \in [0,20]$.} 
\end{figure}

In Fig. \ref{fig:fig2} (a) we show a plot of $M_{LR}$ as a function of the coupling strength $\gamma_0$, calculated both with and without the RWA. As was observed in the static regime, the LR measure presents a similar behaviour in both cases. In particular, it can be seen that $M_{LR}$ takes values of the order of 10 for small $\gamma_0$, and then decays to its minimum possible value of 1 as the coupling increases. Physically, this tells us that the driving is able to increase the non-Markovianity of the evolution well above the value of the static case, but can only do so in the weak coupling regime. For strong coupling, on the other hand, the driving fails to reach values of NM significantly above the static ones. This is the main result of our work. We stress that the large enhancement of the NM values appear because the revivals of the distinguishability in the static case are small. Nevertheless, it is surprising that the influence of the driving field is much stronger in the weak-coupling case, which is precisely where non-Markovian effects are usually neglected or discarded by perturbative approaches.\\

In Fig. \ref{fig:fig2} (b) we show results obtained for $M_{BLP}$, the maximum NM relative value computed with the BLP measure. There, it can be seen that calculations made using the rotating wave approximation also describe the phenomenon of NM enhancement at weak coupling. However, in the case where the RWA is not applied, $\Nc_{BLP}$ shows a completely different behaviour, remaining practically constant as the coupling $\gamma_0$ varies. These findings are reminiscent of the analysis we performed for the static case. This provides further proof of the reliability of the LR measure as opposed to the BLP measure.\\

It is interesting to point out that a number of previous works have reported the appearance of non-Markovian effects for small coupling due to driving acting on the system \cite{bib:addis2015,bib:haikka2011,bib:zhang2015,bib:soares2016}). In those cases, however, the driving was introduced as a new interaction term in the Hamiltonian of the system. Here we show that NM can be induced by just allowing the original system parameters to change in time.\\

\section{Discussion} \label{sec:discussion}


\begin{figure}
	\includegraphics[width=\linewidth]{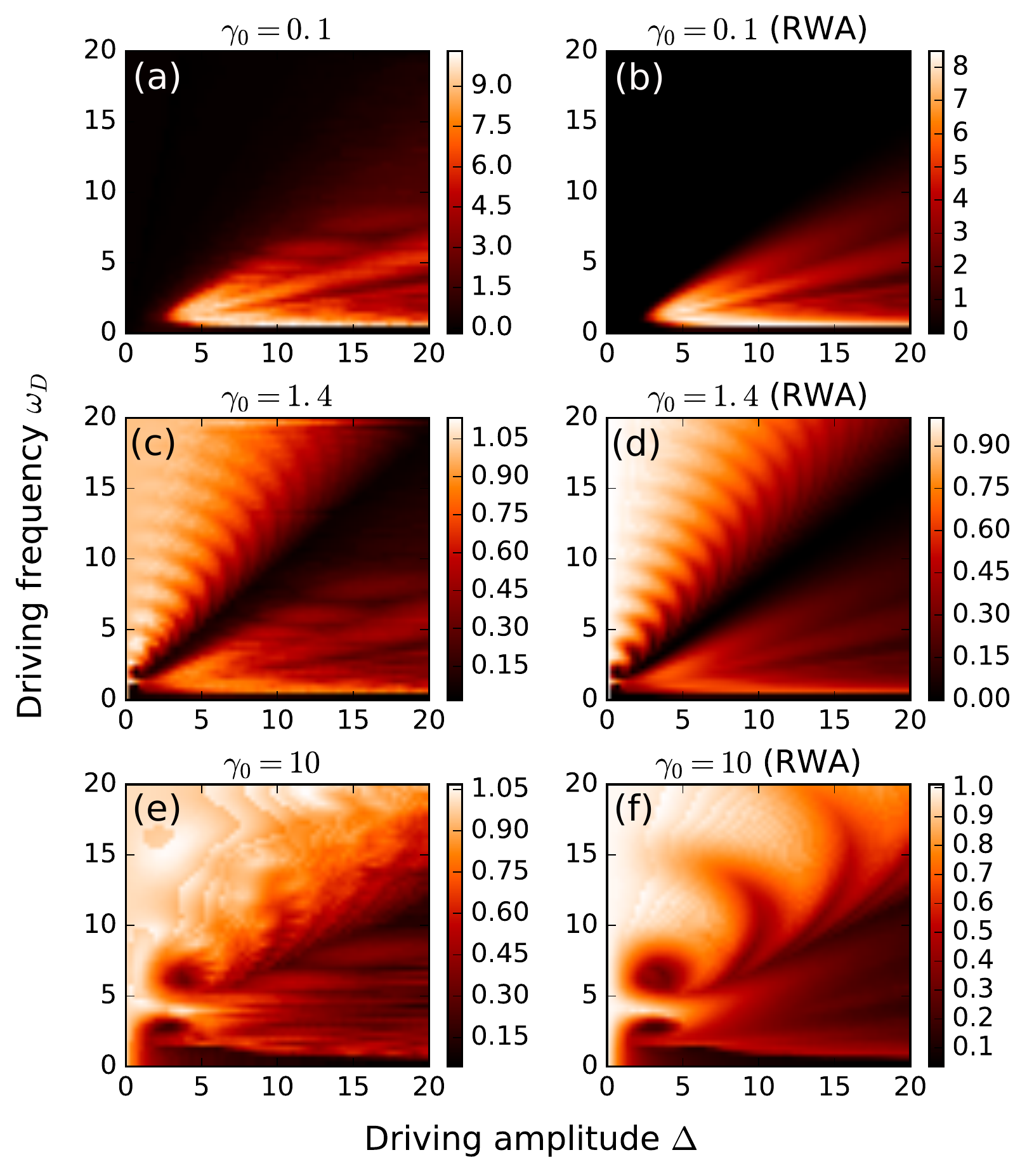}
	\caption{\label{fig:fig3} Density plots for the relative non-Markovianity value $\Nc_{LR}^{rel}$ computed for the largest revival (LR) measure for different values of the coupling strength $\gamma_0$. Results are shown as a function of the dimensionless driving amplitude $\Delta$ and frequency $\omega_D$ in all cases. (a) and (b) $\gamma_0=0.1$. (c) and (d) $\gamma_0=1.4$. (e) and (f) $\gamma_0=10$. Plots in the right column were obtained under using the equation of motion (\ref{ec:ec_rwa}), i.e. using the RWA. Plots in the left column were obtained using the hierarchy equation method (beyond RWA). Dimensionless parameter values: $\Omega_0=20$.}
\end{figure}

Let us explore in further detail the enhancement of non-Markovian effects induced by the driving field. In Fig. \ref{fig:fig3} we show several density plots of the relative LR measure $\Nc_{LR}^{rel}$ for different values of the coupling strength, as a function of the driving frequency $\omega_D$ and amplitude $\Delta$. From the figure, we can roughly separate the behaviour of the NM values for the regions where $\omega_D/\Delta>1$ and $\omega_D/\Delta<1$, which we will referr to as high and low frequency regimes, respectively. In particular, for small coupling (top row in Fig. \ref{fig:fig3}), we observe $\Nc_{LR}^{rel}\simeq0$ for high frequency driving, while for  $\omega_D/\Delta<1$ we observe the enhancement of NM discussed earlier. When the system is strongly coupled to the environment (bottom row in Fig. \ref{fig:fig3}), the situation is inverted, and the largest values of $\Nc_{LR}^{rel}\simeq1$  occurr when $\omega_D/\Delta>1$.  For low frequencies, the relative NM is significantly smaller than 1, meaning that the turning on the driving actually causes distinguishability revivals to decrease during the evolution of the system. In the intermediate coupling regimes, we observe a clear transition between both cases. It can also be seen that there is a large region, corresponding to $\omega_D/\Delta\simeq1$, where NM is suppressed altogether.\\

We will now look closer to the high frequency case $\omega_D/\Delta>1$. There, the driving fails to increase the degree of NM with respect to the static case ($\Nc_{LR}^{rel}\leq 1$), for all coupling strengths. This behaviour can be explained analytically in the following way. By transforming the original Hamiltonian (\ref{ec:hami}) to a rotating frame via the transformation 

\begin{equation}
U_R(t)=\expo{-i\left(\int_0^t \bar{\Delta} \mathrm{cos}(\bar{\omega}_D s) ds\right)\sigma_+\sigma-},
\end{equation}

\noindent it is straightforward to show that, for $\omega_D\gg1$, the dynamics is set by a transformed Hamiltonian:

\begin{eqnarray}
H_R(t)&=&U_R(t)\left[H-\bar{\Delta}\mathrm{cos}(\bar{\omega}_D t)\right]\daga{U}_R(t) \label{ec:hami_r} \\
&\simeq&\bar{\Omega}_0\sigma_+\sigma_- + \beta\sigma_x\sum_k g_k(b_k+b_k^\dagger) + \sum_k\bar{\omega}_k b_k^\dagger b_k, \nonumber 
\end{eqnarray}

\noindent where $\beta=J_0\left(\frac{\bar{\Delta}}{\bar{\omega}_D}\right)=J_0\left(\frac{\Delta}{\omega_D}\right)\leq 1$, and $J_0(x)$ is a Bessel function (see supplementary material for more details). It is easily seen that (\ref{ec:hami_r}) can be interpreted as the original Hamiltonian (\ref{ec:hami}) but with no driving, zero detuning and a with modified coupling constants $g_k\rightarrow\beta g_k$. This, in turn, is equivalent to having a modified coupling parameter in the Lorentzian spectral density (\ref{ec:spectral}) 

\begin{equation}
\gamma_0\rightarrow\beta^2\gamma_0\leq\gamma_0.
\end{equation}

In short, when the driving frequency is high, the  driven system dynamics is equivalent (up to a unitary transformation) to that of a static system which couples less strongly to the environment. As we showed in Fig. \ref{fig:fig1}, the degree of NM in that case increased monotonically with $\gamma_0$. Note also that, since the trace distance (\ref{ec:trace_dist}) is invariant under unitary transformations, the degree of NM is independent of $U_R(t)$.  As a result, the high frequency driving fails to increase the NM measures above the static case, thus proving our numerical finding. It is worth pointing out that this phenomenon was previously discussed in the context of dynamical decoupling techniques \cite{bib:fonseca2004}.\\

\begin{figure}
	\includegraphics[width=\linewidth]{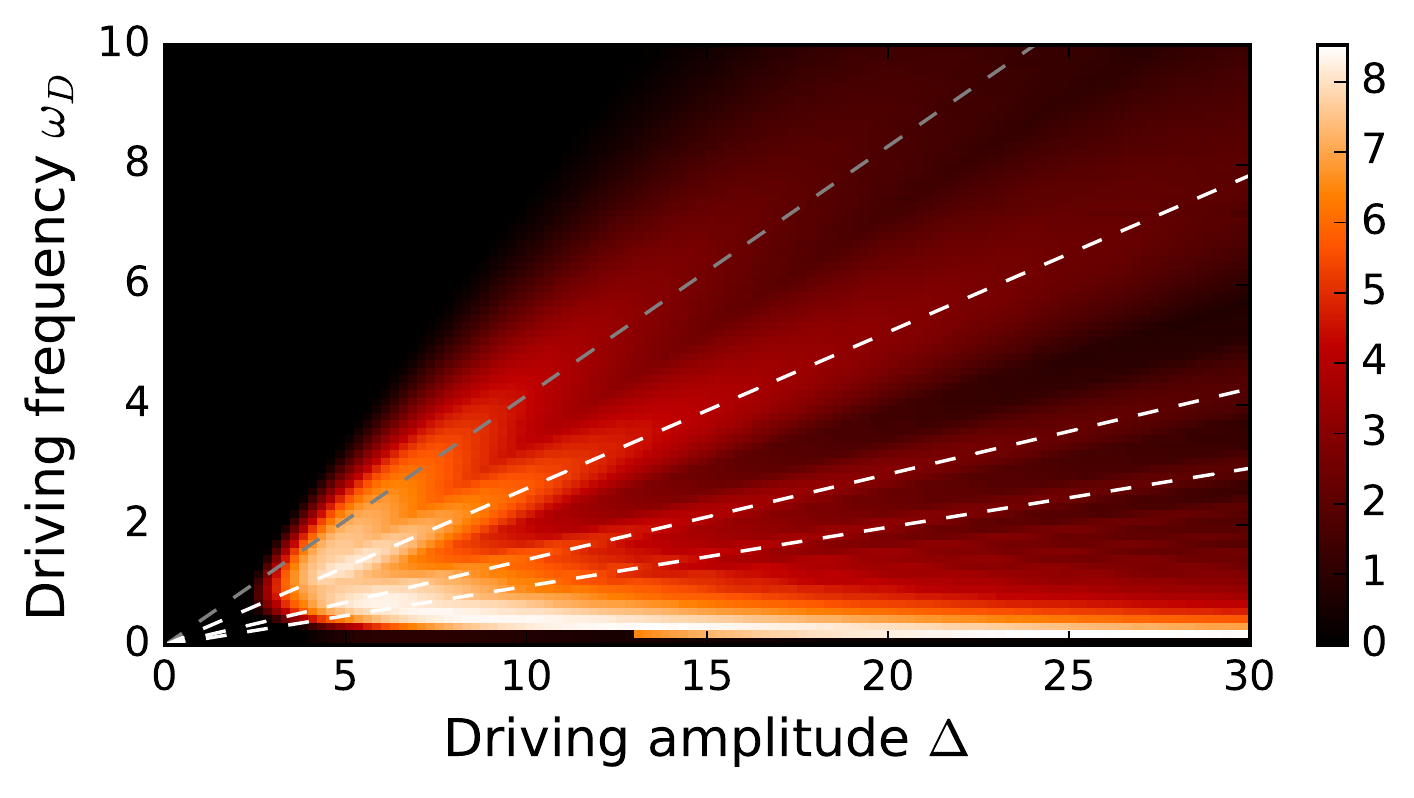}
	\caption{\label{fig:fig4} Density plot for the relative non-Markovianity value $\Nc_{LR}^{rel}$ computed for the largest revival (LR) measure. Parameters as in Fig. \ref{fig:fig3} (b). Dashed lines represents the values of $\omega_D/\Delta$ for which $J_0\left(\frac{\Delta}{\omega_D}\right)=0$ (gray) and $J_1\left(\frac{\Delta}{\omega_D}\right)=0$ (white).}
\end{figure}

In agreement with the previous discussion, for the weak coupling case we observe the largest values of NM in the low frequency regime, see Fig. \ref{fig:fig3} (a) and (b). We show a detailed plot of the NM relative value for this region in Fig. \ref{fig:fig4}. There, it can be seen that maxima for $\Nc_{LR}^{rel}$ spread along straight lines in the $(\Delta,\:\omega_D)$ plane. Interestingly, this lines coincide with the appearance of zeros of  $J_0\left(\frac{\Delta}{\omega_D}\right)$ and $J_1\left(\frac{\Delta}{\omega_D}\right)$. In fact, this behaviour can be explained from the analytical solution for $G(\tau)$ which can be obtained under the RWA to first order in the dimensionless coupling parameter $\gamma_0$. The resulting expression is rather complicated, and it is shown in the supplementary material. It is interesting to point out that parameter regions determined by the Bessel functions are typical in periodically driven quantum systems \cite{bib:grossmann1991,bib:poggi2014}.\\

\section{Concluding remarks} \label{sec:concluding}

In this paper we have explored the interplay between driving and non-Markovianity on the dynamics of an open quantum system. By numerically studying the proposed model for various parameter regimes, we find an remarkable result: the driving can produce a large enhancement non-Markovian effects, but only when the coupling between system and environment is small. In the strong-coupling regime, on the other hand, the driving is unable to increase the degree of NM. We performed extensive numerical calculations which prove that this effect is present also beyond the rotating wave approximation. We also provide analytical arguments which allows us to explain the effects of the driving in the high-frequency regime, and also in the weak coupling regime.\\

We believe that the results presented here shed new light on the relation beteween non-Markovianity and controlled open quantum dynamics. While the debate about whether NM could be harnessed as a useful resource is still open, this work provides a first step in the study of how to control such effects with external time-dependent fields. \\

\begin{acknowledgments}
We acknowledge support from CONICET, UBACyT, and ANPCyT (Argentina).
\end{acknowledgments}

\bibliography{nm_arxiv.bib}


\onecolumngrid
\section{Supplementary Material}

\subsection{Hierarchy equation method}

Consider the total system Hamiltonian which describes the dynamics of the two-level system and the environment

\begin{equation}
H = \omega_0(t)\sigma_+\sigma_- + \sigma_x\sum_k g_k(b_k+b_k^\dagger) + \sum_k\omega_k b_k b_k^\dagger.
\label{ec:hami}
\end{equation}

From Eq. (\ref{ec:hami}), it is straightforward to write down a formal solution for the reduced density matrix of the two-level system in the interaction picture
\begin{equation}
\tilde{\rho}(t)=\mathcal{T}\mathrm{exp}\Bigg\{-\int_0^t dt_2\int_0^{t_2}dt_1\: \tilde{\sigma}_x(t_2)^\times\Big[C_R(t_2-t_1)\tilde{\sigma}_x(t_1)^\times
+ i\:C_I(t_2-t_1)\tilde{\sigma}_x(t_1)^\circ\Big]\Bigg\}\tilde{\rho}(0),
\label{ec:sol_formal}
\end{equation}

\noindent where $\tilde{O}$ denotes an operator in the interaction picture and $\mathcal{T}$ is the usual time-ordering operator. Functions $C_R(t)$ and $C_I(t)$ are the real and imaginary parts of the baths correlation function, which is defined as
\begin{equation}
C(t-t')=\mathrm{tr}\left[\rho_B B(t)B(t')\right]=\int_0^{\infty} d\omega\: J(\omega)\ex{-i\omega(t-t')}.
\label{ec:correlat}
\end{equation}

In (\ref{ec:sol_formal}) we have also used a short-hand notation for the commutator and anticommutator operations: $A^\circ B=\left\{A,B\right\}$ and $A^\times B=\left[A,B\right]$. The hierarchy equation method casts the expression (\ref{ec:sol_formal}) into a set of ordinary differential equations
\begin{equation}
\frac{d \varrho_{\vec{n}}}{d\tau}=-\left[i H_s(\tau)^\times+\vec{n}.\vec{\nu}\right]\varrho_{\vec{n}}-i\sum_{k=1}^{2}\Big\{\sigma_x^\times\varrho_{\vec{n}+\vec{e}_k}+\frac{\gamma_0}{2}n_k\left[\sigma_x^\times+(-1)^k \sigma_x^\circ\right]\varrho_{\vec{n}-\vec{e}_k}\Big\},
\label{ec:hierar}
\end{equation}

\noindent where we have defined $\vec{\nu}=(1-i\Omega_0,1+i\Omega_0)$ and $\vec{n}=(n_x,n_y)$, where $n_x,n_y\geq0$. Note that equations (\ref{ec:hierar}) have been expressed in dimensionless units as explained in the main text. This system can be solved for the operator $\varrho_{\vec{n}}(\tau)$, with initial conditions

\begin{equation}
\varrho_{\vec{n}}(0)=\Bigg\{ \begin{array}{c l }
\rho(0) & \mathrm{for}\ \vec{n}=(0,0) \\
0 & \mathrm{otherwise} \end{array}
\end{equation} 

Of course, the hierarchy equations  (\ref{ec:hierar}) must be truncated for large enough $\vec{n}$ for numerical purposes. In this work we take $n_x,n_y\leq N$. Using $N$ of the order of 10 gives a converged, positive density matrix $\rho(\tau)$.

\section{High frequency equivalent Hamiltonian}

To analyze the dynamics generated by Hamiltonian (\ref{ec:hami}), let us turn to a rotating frame via the unitary transformation
\begin{equation}
U_R(t)=\expo{-i\left(\int_0^\tau \Delta \mathrm{cos}(\omega_D s) ds\right)\sigma_+\sigma-}
\end{equation}

The complete (system plus bath) state in this frame $\Ket{\Psi_R(t)}$ is related to rest frame representation via $\Ket{\Psi_R(t)}=U_R(t)\Ket{\Psi(t)}$. The equation for the evolution of $\Ket{\Psi_R(t)}$ can be expressed in the following way
\begin{equation}
\frac{d}{dt}\Ket{\Psi_R(t)}=i\:H_R(t)\Ket{\Psi_R(t)},
\end{equation}

\noindent where $H_R(t)$ is the transformed Hamiltonian 
\begin{equation}
H_R(t)=U_R(t)\left[H(t)-\Delta\mathrm{cos}(\omega_D t)\right]\daga{U_R}(t)=
\Omega_0\sigma_+\sigma_- + \sigma_x^R(t)\sum_k g_k(b_k+b_k^\dagger) + \sum_k\omega_k b_k b_k^\dagger,
\label{ec:hami_r}
\end{equation}

\noindent where $\sigma_x^R(t)=U_R(t)\sigma_x\daga{U_R}(t)$. By using the following identity
\begin{equation}
\ex{i \gamma\:\mathrm{sin}(\Omega t)}=\sum\limits_{n=-\infty}^{+\infty}J_n(\gamma)\ex{in\Omega t},
\label{ec:expansion}
\end{equation}

\noindent we can express $\sigma_x^R(t)$ in the following way
\begin{equation}
\sigma_x^R(t)=\mathrm{cos}(\alpha(t))\sigma_x+\mathrm{sin}(\alpha(t))\sigma_y
\label{ec:sigmax_r}
\end{equation}

\noindent where
\begin{equation}
\alpha(t)=\frac{\Delta}{\omega_D}\mathrm{sin}(\omega_D t).
\end{equation}

We can now take the high-frequency limit by assuming that $\omega_D/\lambda\gg1$. In such case, we can neglect the fast-oscillating terms in (\ref{ec:expansion}) and keep only the constant contribution, such Eq. (\ref{ec:sigmax_r}) now takes the form
\begin{equation}
\sigma_x^R(t)\simeq J_0\left(\frac{\Delta}{\omega_D}\right)\sigma_x,
\end{equation}

\noindent and thus the transformed Hamiltonian (\ref{ec:hami_r}) is time-independent. 

\section{Weak coupling solution}

Consider the equation of motion for $G(\tau)$ under the rotating wave approximation (RWA) as introduced in the main text
\begin{equation}
G''(\tau)+\left[1-i\Delta\mathrm{cos}\left(\omega_D\tau\right)\right]G'(\tau)+\frac{\gamma_0}{2}G(\tau)=0,
\label{ec:ec_rwa}
\end{equation}

\noindent where $G(0)=1$ and $G'(0)=0$. We now treat the term $\frac{\gamma_0}{2}G(\tau)$ as a perturbation and expand the solution in powers of $\gamma_0$
\begin{equation}
G(\tau)=\sum_n \gamma_0^n g_n(\tau).
\label{ec:pert}
\end{equation}

By inserting (\ref{ec:pert}) in (\ref{ec:ec_rwa}) we obtain 
\begin{equation}
\sum_n\gamma_0^n g_n''(\tau)+[1-i\Delta(\tau)]\sum_n\gamma_0^n g_n'(\tau)+\frac{1}{2}\sum_n\gamma_0^{n+1} g_n(\tau)=0,
\end{equation}

\noindent where we have used the short-hand notation $\Delta(\tau)\equiv\Delta\mathrm{cos}(\omega_D\tau)$. By arranging powers of $\gamma_0$ we obtain the equation for the zeroth-order term $g_0(\tau)$,

\begin{equation}
\Bigg\{\begin{array}{r c l}
g_0''+[1-i\Delta(\tau)]g_0'&=&0\\
g_0(0)&=&1\\
g_0'(0)&=&0,\end{array},
\end{equation}

\noindent which has the trivial solution $g_0(\tau)=1$. For the following order, we obtain
\begin{equation}
\Bigg\{\begin{array}{r c l}
g_1''+[1-i\Delta(\tau)]g_1'+\frac{1}{2}g_1&=&0\\
g_1(0)&=&0\\
g_1'(0)&=&0,\end{array}.
\end{equation}

This system can be integrated in a straightforward fashion to obtain
\begin{equation}
g_1'(\tau)=-\frac{1}{2}\sum_n\sum_m \frac{\jota{n}\jota{m}}{1-in\omega_D}\left(\ex{-i(n-m)\omega_D\tau}-\ex{-\tau}\ex{im\omega_D \tau}\right)
\label{ec:g1}
\end{equation}

We can obtain $g_1(\tau)$ by integrating expression (\ref{ec:g1}). The full approximated solution is then given by
\begin{equation}
G(\tau)\simeq g_0(\tau)+\gamma_0 g_1(\tau)=1+\gamma_0 g_1(\tau).
\end{equation}

\end{document}